\documentclass[prb,reprint,amsmath,amssymb,showpacs,superscriptaddress,floatfix,longbibliography]{revtex4-2}
\usepackage[breaklinks=true,colorlinks,citecolor=blue,linkcolor=blue,urlcolor=blue]{hyperref}
\usepackage{epsfig,mathrsfs,color,latexsym,subfigure,marginnote,gensymb,}
\usepackage{graphicx}
\graphicspath{{Figures/}}
\usepackage{xcolor}
\usepackage{chemformula}
\renewcommand{\BibitemShut}[1]{}

\usepackage[utf8]{inputenc}

\begin{document}
\title{Strain-tunable in-plane ferroelectricity and lateral tunnel junction in monolayer
	group-IV monochalcogenides}
\author{Achintya Priydarshi}
\affiliation{Department of Electrical Engineering, Indian Institute of Technology, Kanpur, Kanpur 208016, India}
\author{Yogesh Singh Chauhan}
\affiliation{Department of Electrical Engineering, Indian Institute of Technology, Kanpur, Kanpur 208016, India}
\author{Somnath Bhowmick}
\email{bsomnath@iitk.ac.in}
\affiliation{Department of Materials Science and Engineering, Indian Institute of Technology, Kanpur, Kanpur 208016, India}
\author{Amit Agarwal}
\email{amitag@iitk.ac.in}
\affiliation{Department of Physics, Indian Institute of Technology, Kanpur, Kanpur 208016, India}

\date{\today}
\begin{abstract} 
2D Ferroelectric materials are promising for designing low-dimensional memory devices. Here, we explore strain tunable ferroelectric properties of group-IV monochalcogenides MX (M=Ge, Sn; X=S, Se) and their potential application in lateral field tunnel junction devices. We find that these monolayers have in-plane ferroelectricity, with their ferroelectric parameters being on par with other known 2D ferroelectric materials. Amongst SnSe, SnS, GeSe, and GeS, we find that GeS has the best ferroelectric parameters for device applications, which can be improved further by applying uniaxial tensile strain. We use the calculated ferroelectric properties of these materials to study the tunneling electroresistance (TER) of a 4 nm device based on lateral ferroelectric tunnel junction. We find a substantial TER ratio $~10^3-10^5$ in the devices based on these materials, which can be further improved up to a factor of 40 on the application of tensile strain.

\end{abstract}
\maketitle
\section{INTRODUCTION}
Conventional ferroelectrics (e.g., BaTiO$_3$ and PbTiO$_3$), possess switchable spontaneous electric polarization and large dielectric permittivity \cite{Burns1973,WEMPLE19681797,Merz}. They are compatible with silicon-based devices \cite{McKee1998} and offer intriguing functionality such as a giant tunneling electroresistance (TER) and negative capacitance (NC). Due to this,  they have attracted significant interest in various applications such as random access memories (RAM),  field-effect transistors (NCFET), sensors, and solar cells \cite{Zhuravlev2005prl,WangNC,Yan,asifNC,Zhigang,hoffmann2019unveiling}. For exploring the possibility of ultralow-power, high-speed, and nanoscale memory devices in reduced dimensions, it is essential to retain the ferroelectricity of materials at the nanoscale and in reduced dimensions. 

As a solution to these limitations, several two-dimensional ferroelectric materials (2D FEs) have been shown to display ferroelectric behaviour even in the monolayer limit of a few \AA, and are attracting significant attention \cite{Shirodkar2014,osada2019rise,qi2021review,chandrasekaran2017ferroelectricity,cui2018two,shang2021two,liu20212d}. Some examples include SnTe \cite{Chang274}, In$_2$Se$_3$ \cite{soleimani2020ferroelectricity,PhysRevLett.125.047601,PhysRevLett.126.057601}, group-V binary compounds \cite{Sante2015}, group-IV tellurides \cite{Wan2017} and elemental group V monolayer \cite{Xiao2018}. An added advantage in 2D materials is the large strain-induced tunability of their electronic \cite{Peng,fei2014strain,achintyaPRB} and ferroelectric properties \cite{haeni2004room,Liu2018}. However, several of these 2D FEs have relatively small spontaneous polarization and are not useful for device applications.

Recently, robust ferroelectricity with transition temperatures higher than the room temperature \cite{Fei-2016} has been found in Group-IV monochalcogenides MX (M=Ge, Sn; X=S, Se) thin films and dynamically stable MXs monolayers (ML) \cite{li2013,Hu_2019,qin2016diverse}. These MX thin films and ML also have a high thermoelectric performance, and giant piezoelectricity \cite{hu2017high,Gomes2015}. The in-plane ferroelectricity in these materials originates from the anharmonicity in their ionic potential. Moreover, the in-plane spontaneous polarization ($P_s$) is along the armchair direction and has a relatively larger value ($P_{s}=1.56-4.88\times 10^{-10}$ C/m) compared to other 2D FEs. These 2D FEs can be used to make a novel completely lateral in-plane ferroelectric tunnel junction (2D FTJ) based on p-type semiconductor/FE/n-type semiconductor structure \cite{shen2019two}. These lateral 2D FTJ can display a giant TER of 1460$\%$. Additionally, vertical 2D FE homojunction-based tunnel field-effect transistor (TFET) has been proposed to have low leakage current ($\sim$5-7 $\mu A/\mu m$) for potential application in low power devices \cite{D0RA03279D}.
\begin{figure*}    
\includegraphics[width=0.9\linewidth]{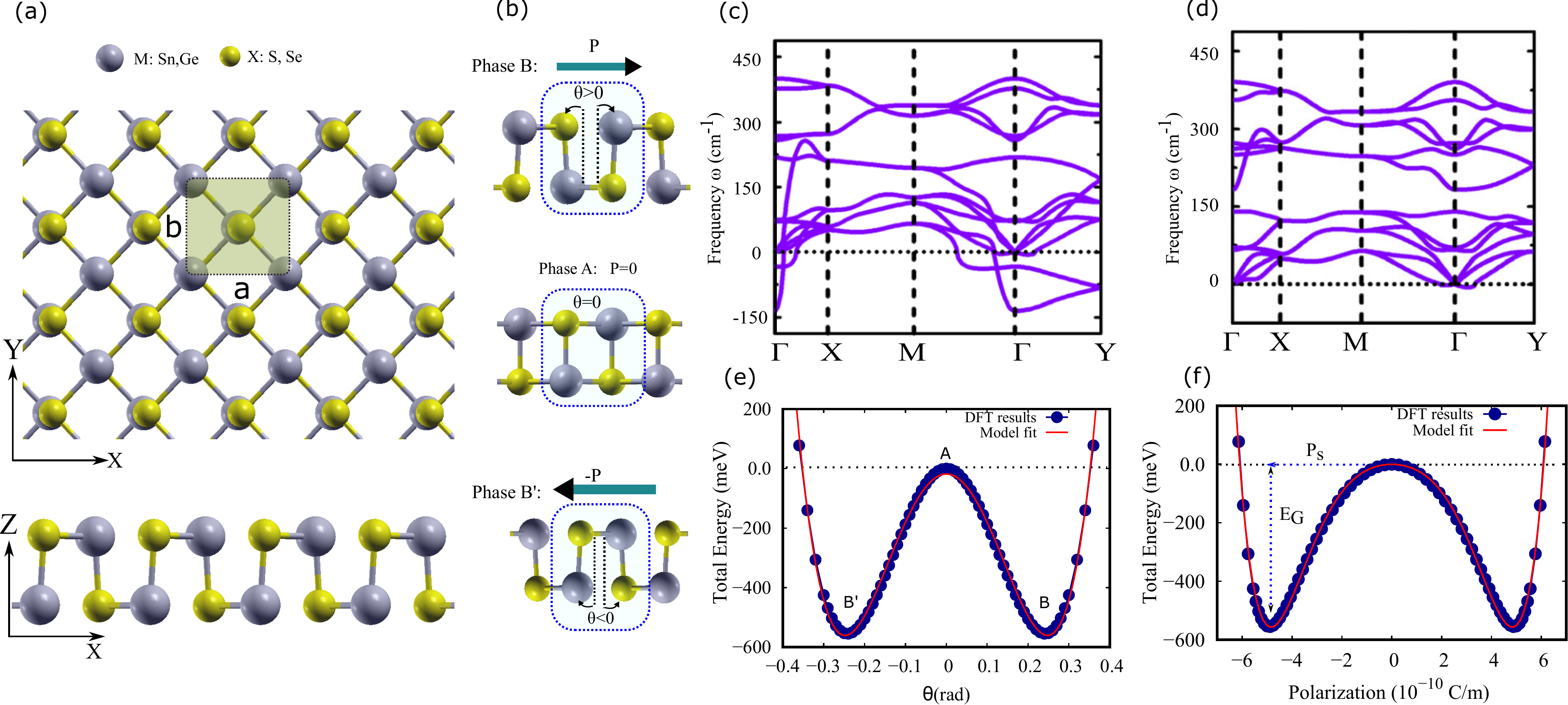}
\caption{(a) Top view and side view of MX monolayer crystal. The shaded rectangular region marks the unit cell, with $a$ and $b$ being the in-plane lattice constants along the armchair and zigzag direction, respectively. (b) Side view of the two degenerate non-centrosymmetric structures (ferroelectric phase $B$ and $B'$) and the centrosymmetric structure (paraelectric phase $A$). Phonon  dispersion of of monolayer GeS structures for the (c) A (unpolarized) phase and (d) B (polarized) phase. (e) Double-well potential as a function of the angular distortion ($\theta$) of monolayer GeS. (f) Double-well potential as a function of the polarization of monolayer GeS. The spontaneous polarization and potential barrier are labeled as $P_s$ and $E_G$, respectively. The fitting curve (red line) is based on the phenomenological Landau model.}
\label{fig:structure}
\end{figure*}

Motivated by these studies, here we explore strain tunability of in-plane ferroelectricity in group-IV monochalcogenides MX (M=Ge, Sn; X=S, Se) and lateral 2D FTJ devices based on them. We find that the value of spontaneous polarization and other ferroelectric parameters for these four MLs are relatively larger than most of the other known 2D materials, and the application of uniaxial strain can further enhance these. We also explore the performance of lateral FTJ based on these monolayers and show that GeSe and GeS display giant TER ($~10^3-10^5$), with GeS having the largest value. The calculated TER value can be further increased by a factor of 40 under the application of $4\%$ uniaxial tensile strain. Our study shows that these MX monolayers can be promising materials for 2D nonvolatile nanoscale devices such as lateral 2D FTJ. 

The organization of the paper is as follows. Section II describes the crystal structure and computational details, and we discuss the in-plane ferroelectricity in these MX monolayers in Sec.~III. The strain engineering of the ferroelectricity is discussed in Sec. IV, and the lateral field tunnel junction device and tunneling electroresistance are described in Sec. V. Our findings are summarized in Sec. VI.
\section{Computational details and Crystal structure}
\label{sec3}

The \textit{ab initio} calculations are performed by density functional theory (DFT), using the projector-augmented wave pseudopotentials and a plane-wave basis set, as implemented in the Quantum Espresso package \cite{Giannozzi_2009}. We treat electron exchange and correlation effects within the generalized gradient approximation (GGA) framework, proposed by the Perdew-Burke-Ernzerhof (PBE) \cite{Perdew1981,PhysRevB.50.17953}. We use the Grimme-D2 \cite{Grimme} method to deal with dispersion forces. We set the kinetic energy cutoff to be 50 Ry for the plane-wave basis set and use a $k-$mesh of $14\times 14\times 1$ for the 2D Brillouin zone (BZ) integrations. We add a vacuum layer of 25{~\AA} along the $z-$direction (perpendicular to the plane of the monolayer) to eliminate any interaction among the replica images. We carry out structural optimizations until all three components of the force on each atom are less than 0.01 eV/{\AA}. We use XCRYSDEN \cite{KOKALJ2003155} for preparing the crystal structures. We adopt the Berry phase method \cite{berryphase} to calculate the spontaneous ferroelectric polarization ($P_{s}$).

A representative crystal structure of the monolayer MX family (M=Sn, Ge; X=S, Se) is shown in Fig.~\ref{fig:structure}(a). For SnSe, we have a rectangular unit cell with $a=4.31$~\AA, $b=4.24$ \AA~, and it has four atoms per cell. Lattice parameters for other materials are reported in the Supporting Information [{See Table S1}]. As shown in Fig.~\ref{fig:structure}(b), the para-electric (PE) phase (space group $Pmmn$) has an inversion symmetry (IS). The IS can be lifted by taking the Sn-Se bond along the $z-$direction and tilting it by an angle $\pm\theta$, leading to the ferroelectric (FE) phase (space group $Pnm2_1$). Two such non-centrosymmetric structures are possible, namely the $B(\theta>0)$ and $B'(\theta<0)$ structures. Each structure can be continuously transformed to the other by spatial inversion [Fig.~\ref{fig:structure}(b)], with the path passing via centrosymmetric phase $A(\theta=0)$. Both $B$ and $B'$ have a polarization of the same magnitude but with opposite signs. Our calculations further reveal that the resultant polarization is an in-plane polarization, with the polarization axis being parallel to the $x$-direction, as marked in Fig.~\ref{fig:structure}.

\begin{figure}  
	\begin{minipage}[b]{0.45\textwidth}  
		\includegraphics[width=0.9\linewidth]{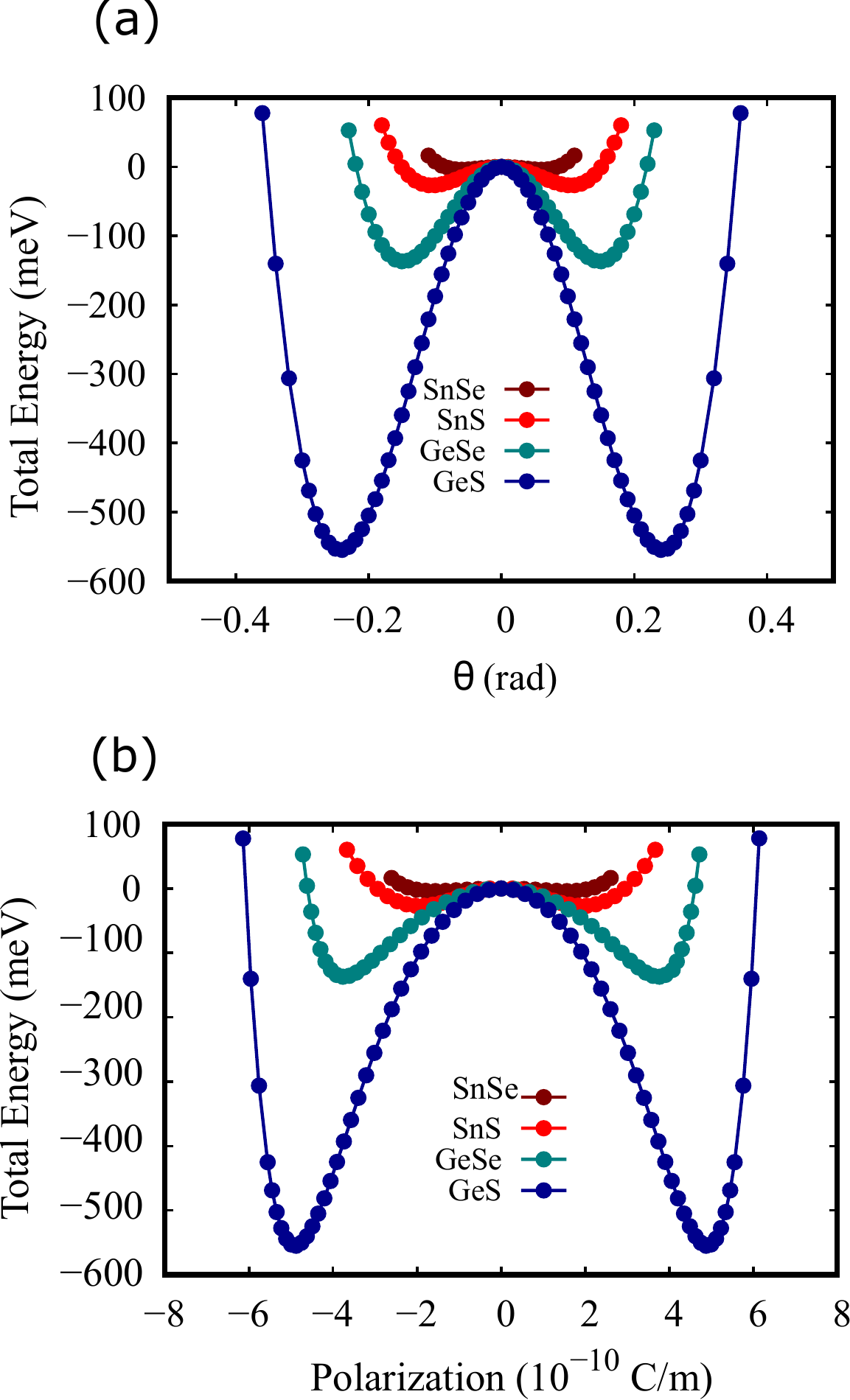}
		\caption{(a) Double-well potential as a function of the angular distortion $\theta$ of monolayer SnSe, SnS, GeSe, and GeS. (b)  Double-well potential as a function of polarization of monolayer SnSe, SnS, GeSe, and GeS. SnSe has the smallest $P_s$ and $E_G$, whereas GeS has the largest $P_s$ and $E_G$ among the studied MX monolayers.}
		\label{fig:w_all_01}
	\end{minipage}
\end{figure}
A comparison of the phonon spectra of PE and FE phase of GeS monolayer is shown in Fig.~\ref{fig:structure}(c) and (d). While the ferroelectric phase has no negative frequencies, the paraelectric phase shows noticeable imaginary frequencies due to the soft optical modes. The existence of such soft phonon modes generally indicate a phase transition from a high symmetry centrosymmetric PE state to a low symmetry non-centrosymmetric FE state below some transition temperature $T_c$ \cite{Fleury}. The phonon dispersion of the remaining monolayers are presented in Fig.~S2 of the supplementary material, and they are found to be dynamically stable.

\section{in-plane ferroelectricity in MX}
A unique feature of the spontaneous polarization in a material is the double well structure in the total energy as a function of the polarization. The double well potential for GeS is shown in Fig.~\ref{fig:structure}(e) for energy vs. $\theta$ and in Fig.~\ref{fig:structure}(f) for energy vs. polarization. The value of the spontaneous polarization and depth of the double well potential for GeS are found to be $P_s=4.88\times 10^{-10}$ C/m and $E_G= -554.85$ meV, respectively. Among the four MX monolayers considered here, SnSe (GeS) has the lowest (highest) $P_s$, as well as $E_G$ values. In particular, GeS has $E_G$ 137 times deeper than the corresponding SnSe value [Fig.~\ref{fig:w_all_01} (a)]. Similarly, GeS has $P_s$ 3 times larger than the corresponding SnSe value [Fig.~\ref{fig:w_all_01} (b)]. We also calculated the saturation polarization ($P_{sat}$) of all four monolayers to explore their usefulness in memory applications such as FeFET. Monolayer GeS has the largest $P_{sat}=6.14\times 10^{-10} C/m$, while SnSe has the lowest $P_{sat}=2.61\times 10^{-10} C/m$ among all four MXs. Taking advantage of larger $P_{sat}$, GeS can achieve a wider memory window and enhance the retention time of a FeFET based memory.

\begin{figure}
	\begin{minipage}[b]{0.45\textwidth}
		\includegraphics[width=0.9\linewidth]{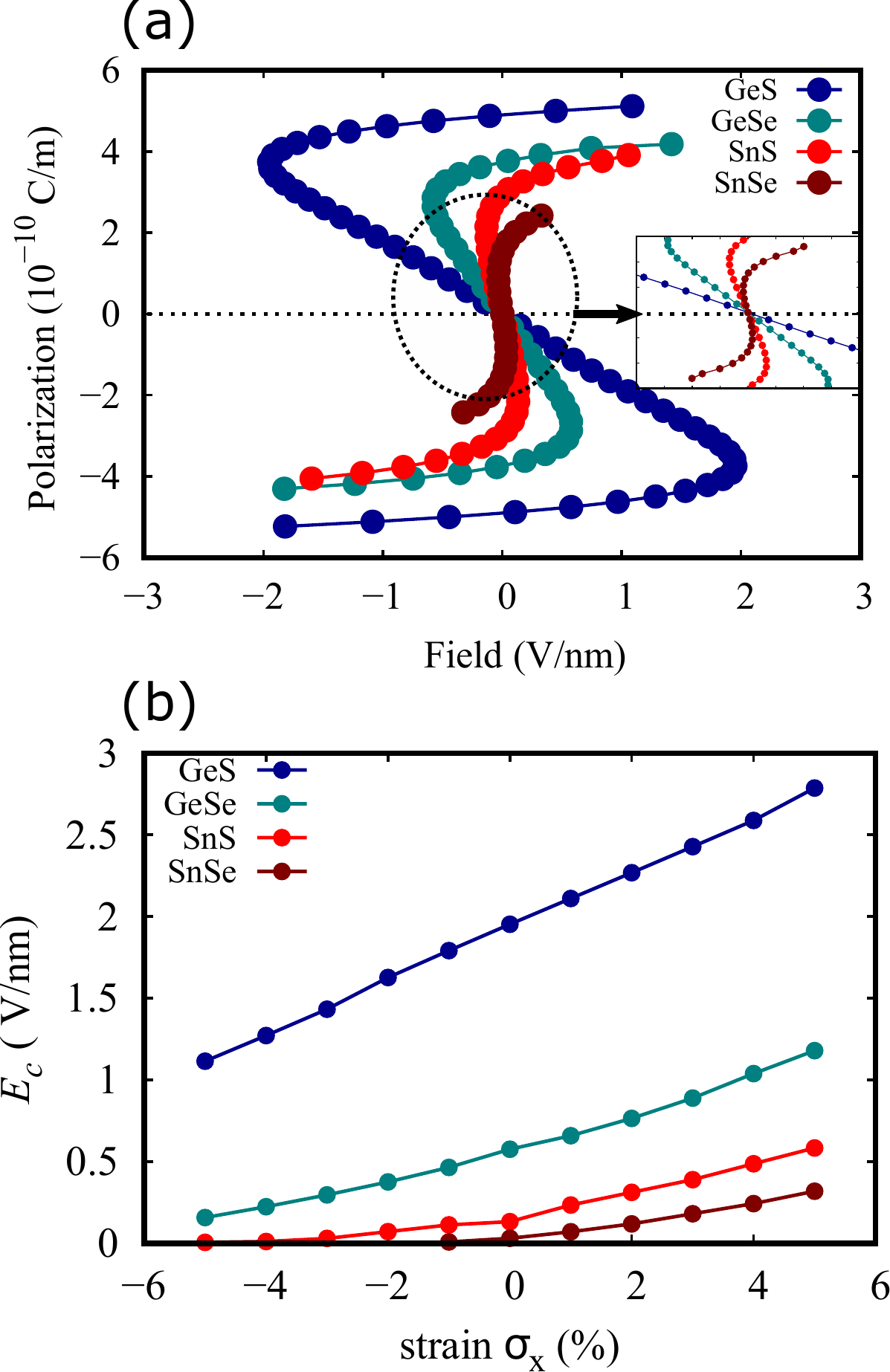}
		\caption{(a) Polarization-field curve (or S-curve) obtained from the data of Fig. \ref{fig:w_all_01}(b). (b) The strain dependence of the in-plane coercive field $E_c$ of MX monolayers, calculated from S-curve. }
		\label{fig:S_Ec}
	\end{minipage}
\end{figure}
\begin{table*}[t]
\caption{The ground-state free energy (potential barrier) $E_G$ (meV), spontaneous polarization $P_{s}$ (10$^{-10}$ C/m), saturation polarization $P_{sat}$ (10$^{-10}$ C/m), and Landau Ginzburg fitting parameters in Eq.~\eqref{eq:1}, $A$, $B$, and $C$, which describe the double-well potential and the coercive field, $E_c$ (V/nm). GeS has the largest spontaneous polarization. \label{T2}}
\centering
{
\begin{tabular}{c c c c c c c c}
\hline \hline
Material    & $E_G$ & $P_{s}$ & $P_{sat}$ & $A$ & $B$ & $C$  & $E_c$\\
\hline \hline
SnSe  & -4.049 & 1.564 & 2.608 & -5.955 & 1.559 & 0.351 & 0.03\\
SnS & -26.741 & 2.036 &3.666 &-29.390 & 8.766 & 0.344 & 0.13\\
GeSe  & -136.965 & 3.768 & 4.716 & -15.434 &  -4.712 & 0.434 & 0.57\\
GeS & -554.845 & 4.877 & 6.142 &-37.385 & -5.787 & 0.315 & 1.95\\
\hline \hline
\end{tabular} }\\
\end{table*}

To understand the ferroelectricity in these materials better, we model the calculated polarization in these materials using the Landau Ginzburg (LG) theory. The LG model relates free energy $(G)$ to polarization $(P)$ via the relation \cite{Fei-2016,Alastair,Wojde}
\begin{equation} \label{eq:1}
G=\dfrac{A}{2}P^{2}+\dfrac{B}{4}P^{4}+\dfrac{C}{6}P^{6}.
\end{equation}
We express $G$ in units of meV per unit cell, and $(P)$ in units of $10^{-10}$ C/m, respectively. The coefficients $A$, $B$ and $C$ are obtained by fitting Eq.~\ref{eq:1} with the values obtained from \textit{ab initio} calculations [Fig.~\ref{fig:structure}(f)].

Another useful ferroelectric parameter is the coercive field $E_c$, the minimum electric field  value needed to switch the polarization direction. In the steady state, the electric field dependence of the polarization (the ``S" curve) is determined by the equation, $E= dG/dP$. Figure~\ref{fig:S_Ec}(a) shows the calculated $S$-curve of the four MX monolayers under strain-free conditions. The $E_c$ can be  calculated from the turning points of the $S$-curve, or $\frac{dE}{dP}|_{E=E_c}=0$. We find that the $E_c$ of GeS is largest, while that of SnSe is the smallest. Our calculated values of $P_{s}$ and $E_c$ are in good agreement with the previously reported theoretical values \cite{Fei-2016}. The experimental range of reported value of $E_c$ in a monolayer SnSe is $\sim$ 1.4-4.5 $\times 10^{5}$ V/cm \cite{Chang2020}, which is in good agreement with our calculated value of $E_c$ $\sim 3\times 10^{5}$ V/cm. The polarization parameters for all four monolayers are tabulated in Table~\ref{T2}.\\
\indent {In 2D In$_2$Se$_3$, the energy needed for the FE to PE transition is about 850 meV \cite{ding2017prediction}, which is considerably higher than that required in the studied monolayers. Since polarization switching has been demonstrated experimentally in In$_2$Se$_3$ \cite{cui2018intercorrelated,ryu2020empowering}, we expect that the monolayers under consideration will also exhibit a polarization switching mechanism under appropriate experimental conditions.
 
A natural question to ask next is what happens to the ferroelectricity in these MX monolayers with strain, which we are going to discuss in the next section.

\begin{figure*}
	\begin{minipage}[b]{0.9\textwidth}
		\includegraphics[width=\linewidth]{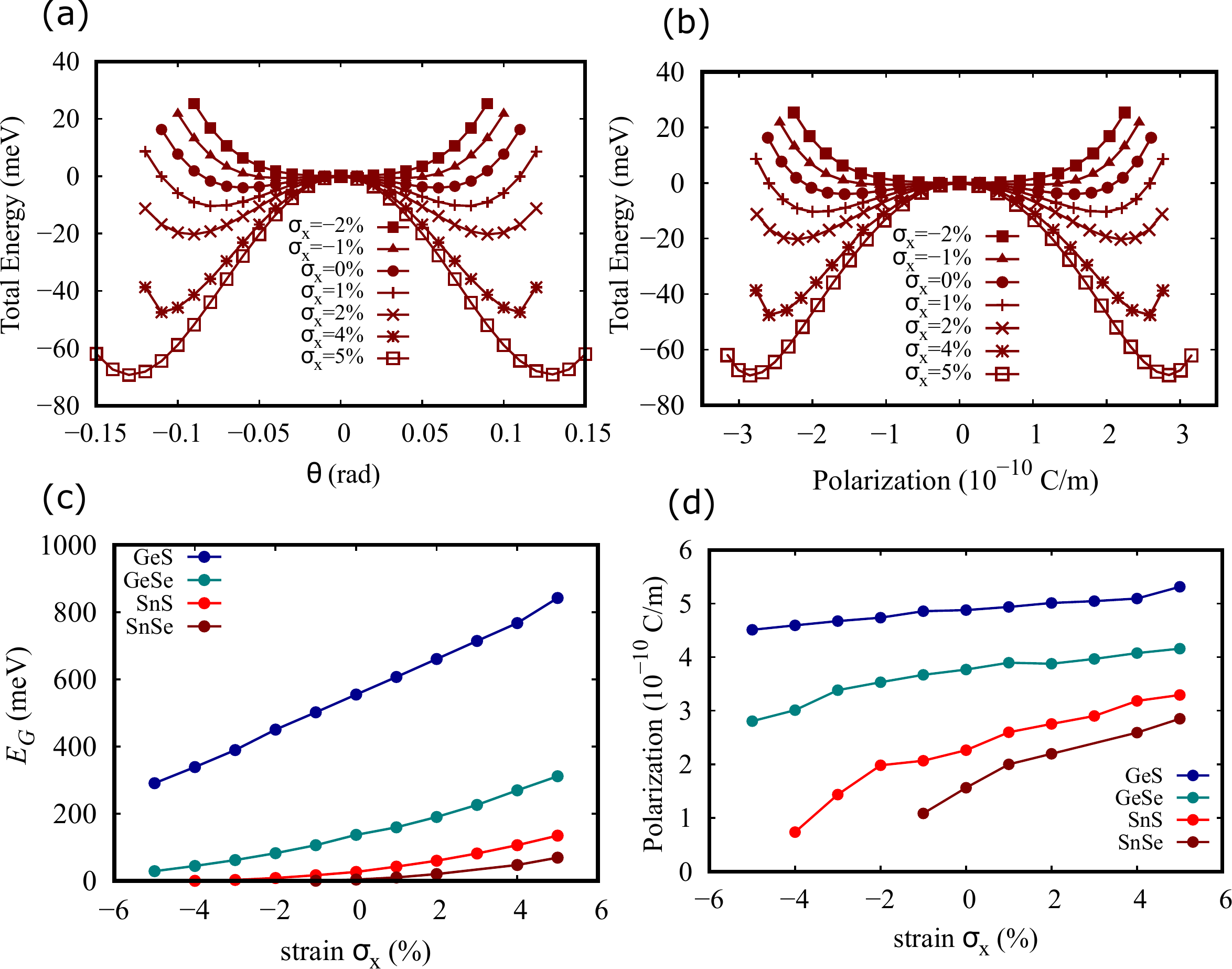}
		\caption{Strain induced modification in the double-well potential as a function of (a) $\theta$, (b) polarization for monolayer SnSe. At $\sigma_{x}=-2\%$, we find that $E_G$ and $P_{s}=0$ for SnSe, and the double well structure in the free energy disappears.
		(c) $E_{G}$ and (d) $P_s$ under the application of compressive and tensile strain.}
		\label{fig:strained_w}
	\end{minipage}
\end{figure*}

\section{Strain engineering of ferroelectricity in MX}

To explore the impact of uniaxial strain on the ferroelectricity, we apply a strain along the polar axis. The strain is introduced by changing the lattice parameter along the $x-$axis: $\sigma_{x}=(\frac{a-a_{0}}{a_{0}})\times100 \% $, where $a$ and $a_{0}$ are the lattice constants along $x$-direction for strained and unstrained unit cells, respectively. Experimentally this uniaxial strain can be achieved by various techniques such as using prestretched substrates, mismatch of thermal expansion, and bending flexible substrate \cite{yang2021strain,li2020efficient,sun2019strain}. The calculated phonon band structures of strained FE phase of considered monolayers are shown in Fig.~S4 of the supplementary information. Importantly, these monolayers are found to be dynamically stable even at two extreme strain conditions. We find that tensile strain along the $x$-direction significantly enhances the  polarization due to incremental change in the in-plane separation between the M (M=Ge,Sn) and the X (X=S,Se) atoms. This leads to an enhancement in the 
$E_G$ and the $P_s$ values. In contrast, compressive strain produces the opposite effect and decreases polarization [see Fig.~S6 in the supplementary material].

The strain induced change in the free energy of SnSe is shown in Fig. \ref{fig:strained_w}(a) and (b) for variation of $\theta$ and the polarization, respectively. As shown in Fig.~\ref{fig:strained_w}(a) and (b), the double well structure in the free energy disappears in SnSe and it becomes a paraelectric material under a relatively small compressive strain of 2\%. On the other hand, tensile strain makes the potential well deeper and increases the value of $P_s$. For example, we find that a 5\% tensile strain enhances $E_G$ by 17 times and $P_s$ by 2 times, compared to the zero strain values of $E_G=-4.05$ meV and $P_s = 1.56\times 10^{-10}$ C/m  [Fig. \ref{fig:strained_w}(c) and (d)].

We find that the other MX monolayers show similar response under strain. However, the magnitude of compressive strain required for FE to PE conversion depends on the depth of the double well potential ($E_G$) of the pristine monolayer. For example, pristine SnS monolayer has a deeper potential ($E_G =-26.74$ meV) compared to pristine SnSe ($E_G =-4.05$ meV). As a result, SnS remains in the FE state up to $\sigma_{x}=-5\%$, while SnSe becomes a PE material at $\sigma_{x}=-2\%$. The variation of $E_G$ and $P_s$ with strain for all four MX monolayers is shown in Fig.~\ref{fig:strained_w}(c) and (d), 
respectively. Since GeSe and GeS have even deeper potential well, having $E_G = -136.97$ meV and $E_G = -554.85$ meV, respectively, they do not undergo FE to PE transition up to compressive strain values of $\sigma_{x}=-5\%$ [Fig. \ref{fig:strained_w}(c)]. 

The strain induced change in the coercive field is shown in Fig.~\ref{fig:S_Ec}(b). We find that the coercive field is very sensitive to strain in all four MX monolayers, and it can change by upto $10$ times for tensile strain upto 5\%. The significant modulation of $E_c$ of the MX monolayers due to applied strain can play a vital role in their potential applications in future electronics. As an example, for compressive strain in SnSe and SnS, 
the $V_c = E_c L < 2$ V (for $L = 20$ nm), which makes them potentially suitable for low power applications.  We now explore the possibility of using the monolayer MX as 2D ferroelectrics tunnel junctions (FTJ).

\section{Lateral Field Tunnel Junction Device and Tunneling Electroresistance}

\begin{figure*} 
	\begin{minipage}[t]{0.9\textwidth}
		\includegraphics[width=0.9\linewidth]{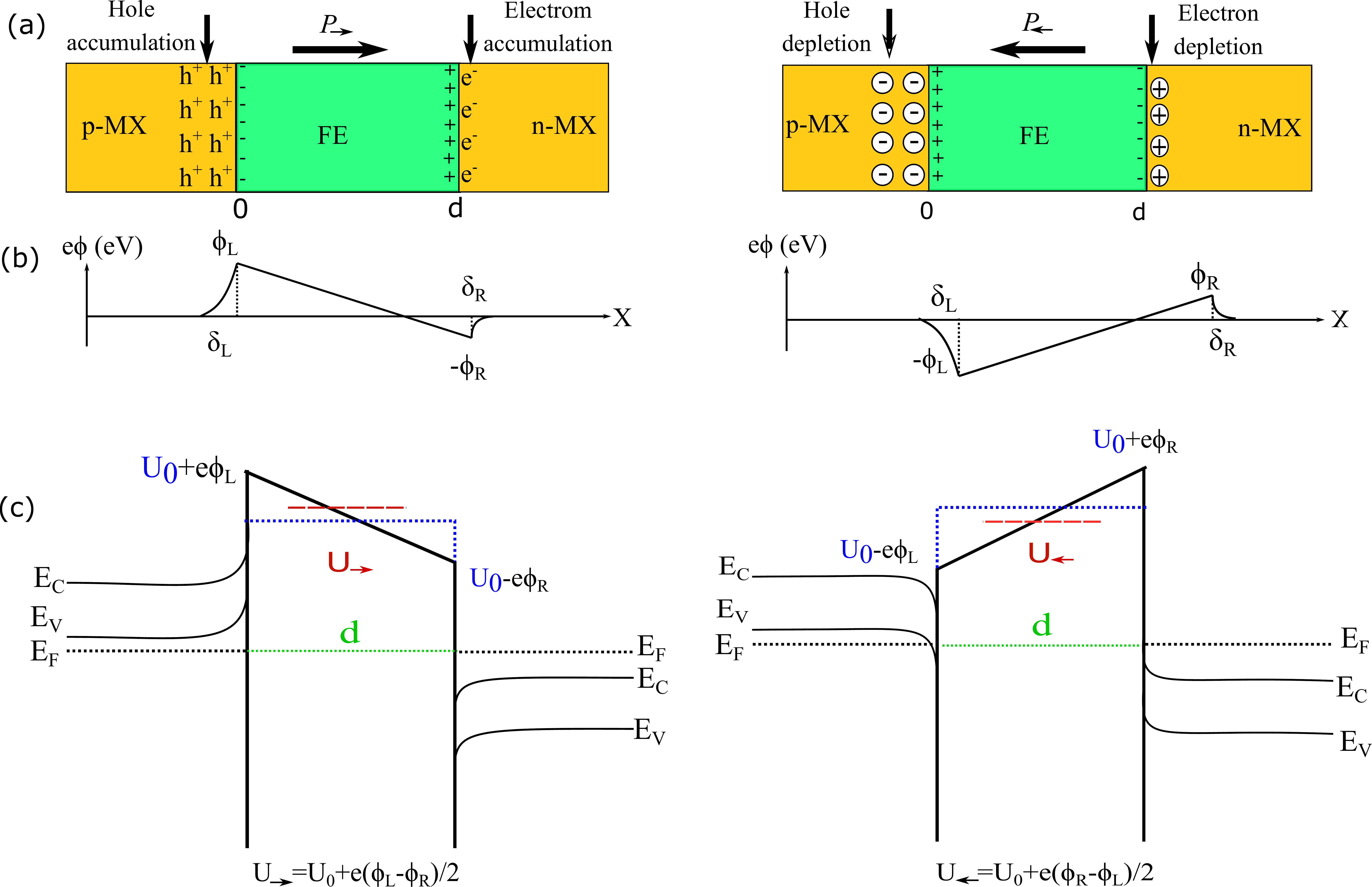}
		\caption{Polarization reversal mechanisms in the 2D-field tunnel junction. (a) Schematic of the device in  P$_{\rightarrow}$ (left panel) and  P$_{\leftarrow}$ (right panel) states. The positive and negative bound charges in the FE region are shown by `+' and `-', respectively. The accumulated electrons and holes in doped regions are represented by `$e^{-}$' and `$h^{+}$'. The symbols `$\oplus$' and `$\ominus$' are used to show the depleted charges in doped regions. (b) Potential energy profile and screening length. (c) Band diagram and potential energy profile of the device. Dotted red line shows the average potential barrier height for two cases, taking $E_{F}$ as a reference. }  
		\label{fig:device_01}
	\end{minipage}
\end{figure*}
The four group IV MX monolayers displaying in-plane ferroelectricity can be used to make a lateral field tunnel junction (FTJ) device \cite{shen2019two} out of the same material. This is in contrast to the generally vertical FTJ which use heterostructures with out of plane ferroelectric materials \cite{garcia2014}. The lateral 2D FTJ device has an added advantage of being relatively easy to fabricate. 

We explore the tunneling characteristics of the MX monolayer based homo-structural lateral FTJ device, as shown in Fig.~\ref{fig:device_01}(a). The electrodes of the device are made of doped MX monochalcogenides (p-type or n-type) whereas as the MX ferroelectric layer is central region. Such a device can be thought to have two tunnel barriers, i) one arising from the mismatched doping in the two electrodes and the device region, and ii) an asymmetric and electrically switchable tunnel barrier arising from the direction of polarization in the ferroelectric material. The combination of these two tunneling barriers gives rise to the tunneling electroresistance (TER) effect \cite{Tsymbal181}.

The tunnel barrier arising from the mismatched doping in the two electrodes and the device part, can be modelled as a rectangular barrier of height $U_{0}$ with respect to the Fermi level $E_{F}$ \cite{Zhuravlev2005prl,Wen2013}. The doping dependent values of $U_{0}$ can be evaluated from first principle calculations, and are listed in Table S2 for all MXs.

\begin{figure} 
	\begin{minipage}[t]{0.45\textwidth}
		\includegraphics[width=0.9\linewidth]{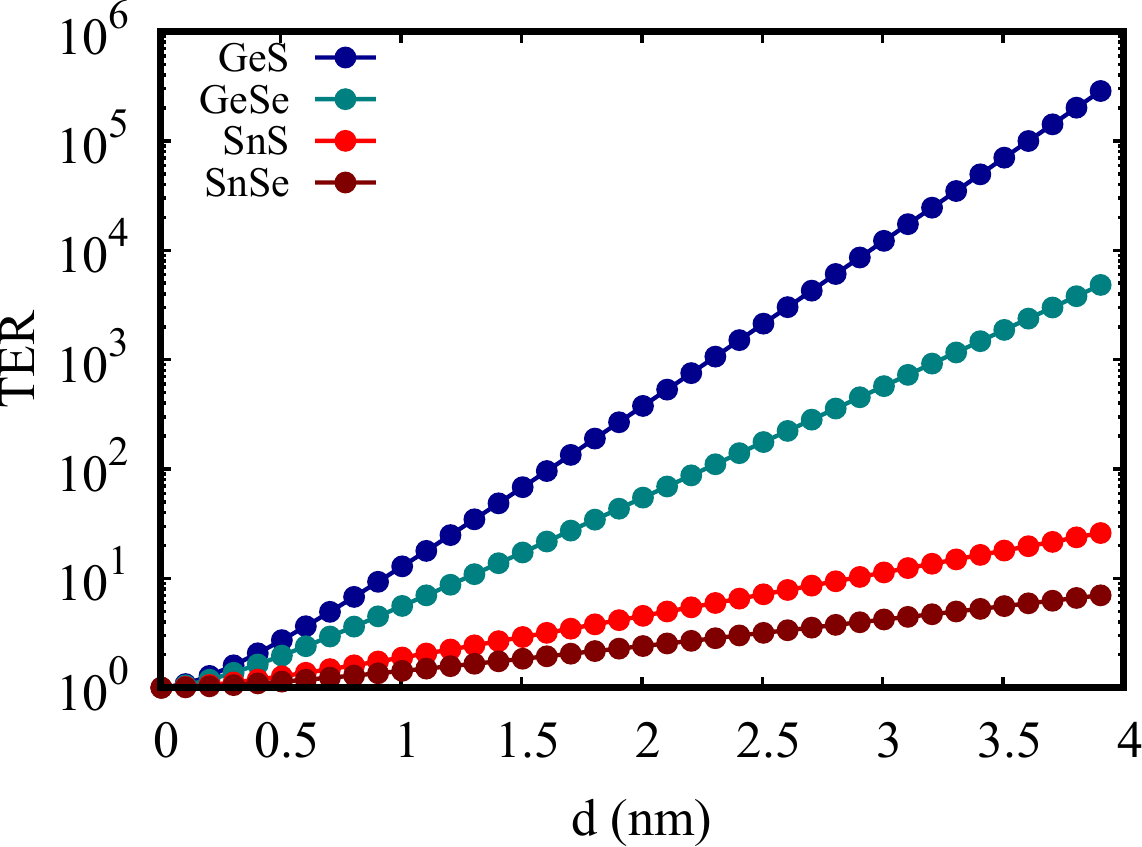}
		\caption{TER as a function of 2D ferroelectric barrier width at zero strain. The parameters used to calculate TER are listed in Table S2.  }
		\label{fig:TER}
	\end{minipage}
\end{figure}
\begin{figure*} 
	\begin{minipage}[t]{0.9\textwidth}
		\includegraphics[width=0.9\linewidth]{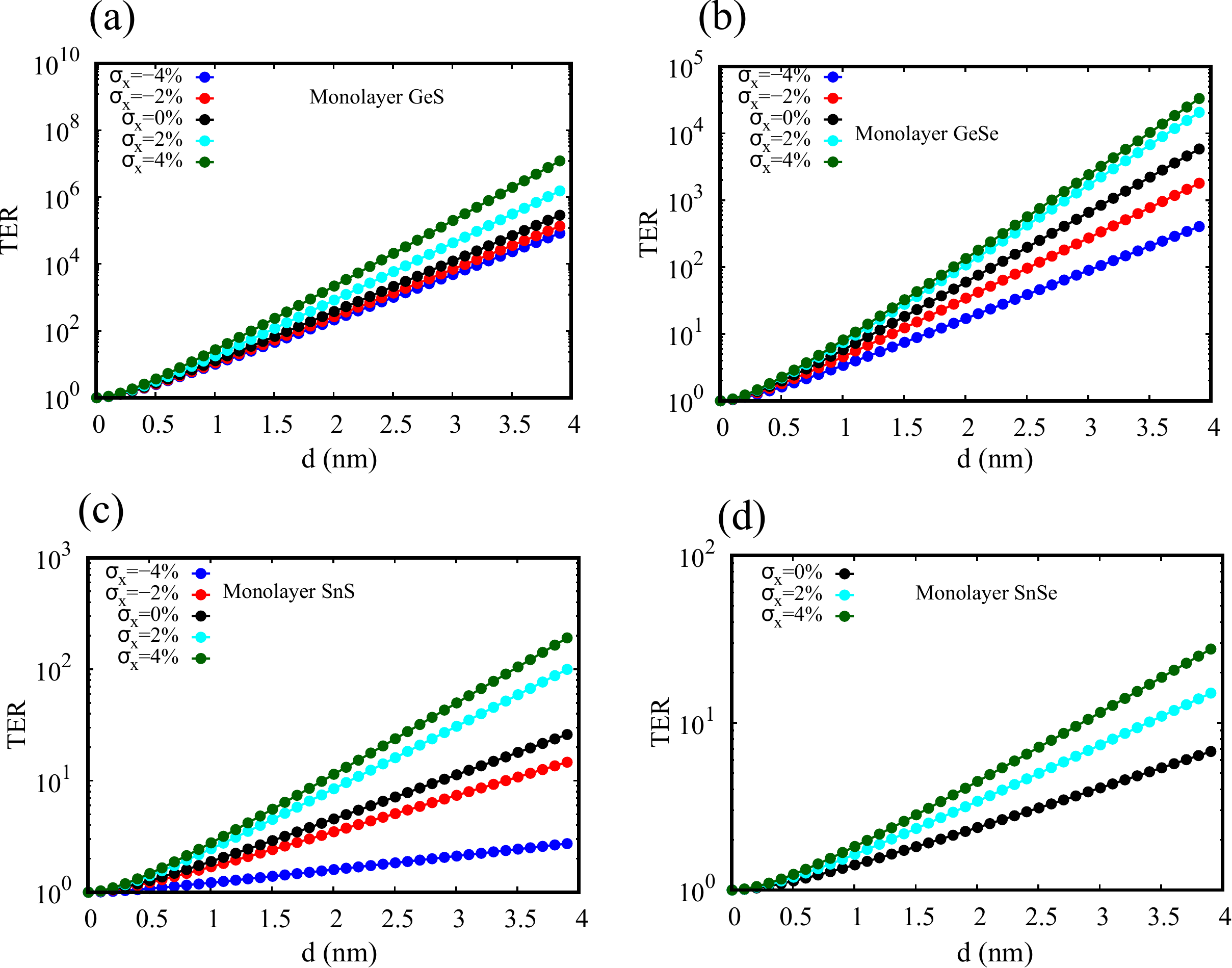}
		\caption{ Strain dependence of TER for (a) GeS, (b) GeSe, (c) SnS and (d) SnSe monolayers. }
		\label{fig:strained_TER}
	\end{minipage}
\end{figure*}

The ferroelectric tunnel barrier arises from the fact that the bound polarization charges at the interface are not completely screened by the adjacent electrodes. This leads to a non zero depolarizing fields in the ferroelectric \cite{Mehta}. The magnitude and shape of the potential associated with this field depends on the magnitude and the direction of the polarization, as shown in Fig.~\ref{fig:device_01}(b). When the polarization is along the positive $x$-direction, (denoted as $P_{\rightarrow}$) we have negative bound charges at the left interface and the positive bound charges at the right interface. Consequently holes are accumulated at the interface in the left electrode and electrons accumulate at the right electrode - see left panel of Fig.~\ref{fig:device_01}(a). For the polarization pointing towards negative $x$-axis ($P_{\leftarrow}$), the situation is shown in the right panel of Fig.~\ref{fig:device_01}(a). This gives rise to an asymmetric potential profile depending on the polarization state of the ferroelectric. 

The magnitude and shape of the of the electrostatic potential at the left and right interface is given by \cite{velev2016npj,Zhuravlev2005prl,shen2019two}
\begin{equation} \label{phi}
	\phi_i=\pm\dfrac{\Gamma_{i}P_sd}{d+\epsilon_{FE}(\Gamma_{L}+\Gamma_{R})}~.
\end{equation}
Here, $i = L/R$, $P_s$ is the induced charge density at the interface of ferroelectric and electrode, and $d$ is the length of the central region, and $\epsilon_{FE}$ is the dielectric constant of the ferroelectric. In Eq.~\eqref{phi}, $\Gamma_{i}=\delta_i/\epsilon_i$ is the ratio of the screening length of the electrode $i$ with its dielectric permittivity $\epsilon_i$. The screening length $\delta$ of a given electrode can be estimated from the Thomas Fermi model. It is given by $\delta=\frac{1}{e}\sqrt{\epsilon/\rho(E_{F})}$ where $\rho(E_{F})$ is the density of states (DOS) at Fermi energy. In our homo-structural device, we have $\epsilon_i = \epsilon_{\rm FE}$. Thus, the screening length of the electrode depends on  $\rho(E_{F})$ only, and it can be tuned by the doping of the electrode. In our device the left (right) electrode is $p$-doped ($n$-doped) and hence we have $\delta_L > \delta_R$, or $\Gamma_L > \Gamma_R$. 

Combining both these potential barriers, we have $U_{\rightarrow}=U_{0}+e(\phi_{L}-\phi_{R})/2$ for the $P_{\rightarrow}$ state and $U_{\leftarrow}=U_{0}+e(\phi_{R}-\phi_{L})/2$ for the $P_{\leftarrow}$ state, respectively \cite{simonsJAP,BrinkmanJAP}. The modulation of the asymmetric trapezoidal tunnel barrier due to the reversal of polarization changes the tunneling conductance, $G$. This gives rise to a finite TER ratio, defined as ${\rm TER}=(G_{\rightarrow}-G_{\leftarrow})/G_{\leftarrow})$, for small applied voltages. A simple estimate for the TER ratio can be done via the Wentzel-Kramers-Brillouin (WKB) approximation for the tunneling probability. This generally works well for $d \sqrt{2m \phi_{1/2}}\gg \hbar$. 

Using the WKB approximation, in the linear response regime, for $\Delta U =U_{\rightarrow}-U_{\leftarrow} \ll U_0$, the TER can be expressed as \cite{gruverman2009,Sokolov_2015} 
\begin{equation} \label{eq:3}
	{\rm TER}\approx \left[ \dfrac{\sqrt{2m}\Delta U}{\hbar \sqrt{U_{0}}}\right] = \exp \left[\dfrac{e}{\hbar}\sqrt{\dfrac{2m}{U_0}} \dfrac{P_{s}\left(\Gamma_{L}-\Gamma_{R} \right)d^{2} }{d+\epsilon_{FE}(\Gamma_{L}+\Gamma_{R})}\right].
\end{equation}
Here, $m$ is the effective mass of the carrier in the barrier region. The TER ratio calculated using Eq.~\eqref{eq:3} is shown in Fig.~\ref{fig:TER} for all four MX monolayers. 
From Fig.~\ref{fig:device_01} (c), it can be clearly seen that the average tunneling barrier in the $P_{\rightarrow}$ state is larger for as compared to $P_{\leftarrow}$ state. 
Thus, the conductance $G_{\rightarrow}$ decays faster than $G_{\leftarrow}$ with  increasing width $d$, causing the TER to increase exponentially with the width. We find that for a $4$ nm device, SnSe and SnS have a TER ratio of $~10$ and $26$, respectively, while the other two MX monolayers (GeS, GeSe) have a relatively large TER $~10^3-10^5$. One reason for this is that ${\rm TER} \propto {\rm exp}[P_s]$, and SnSe has the lowest $P_s$ among all four monolayers. Note that while the TER seems to have an explicit dependence on the effective mass, it is cancelled by the $\Gamma \propto \sqrt{1/\rho(E_F)} \propto 1/\sqrt{m}$ term. We find that for a 4 nm device, the value of TER in GeS and GeSe, is significantly larger than that reported in SnTe based in-plane FJT \cite{ShenFTJ} which has a ${\rm TER}\approx10^2$. Thus, our calculations suggest GeS and GeSe to be good potential candidates for exploring TER based memory devices. 

The variation of $P_s$ with strain (see Fig.~\ref{fig:strained_w}), also manifests in the strain modulation of TER, as shown in Fig.~\ref{fig:strained_TER}.  We find that the ratio of the screening lengths in the two electrodes $\delta_{L}/\delta_{R} \approx 2$ for all four MX monolayers. Further this ratio does not change by much with applied strain. Thus the strain modulation of the TER predominantly arises from the $P_s$. However, the applied strain markedly modulates  $P_s$ and hence TER. We find that under tensile strain of upto $\sigma_x = -4\%$, the TER can increase upto a factor of 40, while a compressive strain of $\sigma_{x}=4\%$ decreases the TER value by 4 times as compared to the unstrained case. 

\section{Conclusion}
\label{sec4}
  In conclusion, we have systematically demonstrated the existence and strain tunability of intrinsic in-plane ferroelectricity in monolayer group-IV monochalcogenides MX (M=Ge, Sn; X=S, Se), using the first-principles DFT calculations. We show that GeS has the largest spontaneous polarization value while SnSe has the lowest value amongst the four MX monolayers. The spontaneous polarization values of GeS are comparable to the largest polarization value reported in other 2D ferroelectric materials such as SbN and BiP \cite{Liu2018}. We find that the tensile strain increases the remnant polarization and $E_G$ in all four monolayers, while the compressive strain decreases it.

Additionally, we demonstrate the applicability of these 4 monolayers in lateral field tunnel junction devices. We find that GeS, and GeSe have a large TER ratio of $10^3-10^5$ making them excellent candidate materials for lateral FTJ devices. Tensile strain of upto $4\%$ can increase the TER in these materials by a factor of 40. Our study will open up further avenues for exploration of fundamental physics and device applications based on the interplay between ferroelectricity and mechanical, electronic and optical properties in these MX monolayers.
\vspace{0.1cm}
\section*{Supplementary Material}
See the supplementary material for (i) the electronic band structure of MX monolayers,  (ii) the dynamical stability of SnSe, GeSe and SnS monolayers, (iii) polar modes in all MX monolayers, (iv) phonon dispersion of monolayers at two extreme strain conditions, (v) strain-induced modification in the double-well potential of GeS, GeSe, and SnS monolayers, (vi) the change in the in-plane separation of MX atoms as a function of strain $\sigma_{x}$, and polarization as a function of in-plane separation, (vii) the strain dependence of S-curve, (viii) the strain dependence of the coercive voltage $V_c$, and (ix) relevant parameters of pristine monochalcogenides used in TER calculation. 
\section*{Acknowledgements}
We acknowledge financial support by the Swarnajayanti Fellowship (Grant No. DST/SJF/ETA-02/2017-18), the FIST Scheme (Grant No. SR/FST/ETII-072/2016) of the Department of Science and Technology, India, the Berkeley Device Modeling Center and the Science and Engineering Research Board, India (Grant No. EMR/2017/004970). We acknowledge National Supercomputing Mission (NSM) for providing computing resources of ``PARAM Sanganak'' at IIT Kanpur, which is implemented by C-DAC and supported by the Ministry of Electronics and Information Technology (MeitY) and Department of Science and Technology (DST), Government of India. We also acknowledge the HPC facility provided by CC, IIT Kanpur.
\section*{Conflict of Interest}
The authors declare no conflict of interest.
\section*{Data Availability Statement}
The data that support the findings of this study are available from the corresponding author upon reasonable request.

\bibliography{ref}
\end{document}


\title{Supporting Information for\\ Strain-tunable in-plane ferroelectricity and lateral tunnel junction in monolayer group-IV monochalcogenides}
	\author{Achintya Priydarshi}
	\affiliation{Department of Electrical Engineering, Indian Institute of Technology, Kanpur, Kanpur 208016, India}
	\author{Yogesh Singh Chauhan}
	\affiliation{Department of Electrical Engineering, Indian Institute of Technology, Kanpur, Kanpur 208016, India}
	\author{Somnath Bhowmick}
	\email{bsomnath@iitk.ac.in}
	\affiliation{Department of Materials Science and Engineering, Indian Institute of Technology, Kanpur, Kanpur 208016, India}
	\author{Amit Agarwal}
	\email{amitag@iitk.ac.in}
	\affiliation{Department of Physics, Indian Institute of Technology, Kanpur, Kanpur 208016, India}
	\date{\today}
	\maketitle
\begin{figure*}[t]
	\centering	
	\begin{minipage}[!h]{0.99\textwidth}
	\includegraphics[width=\linewidth]{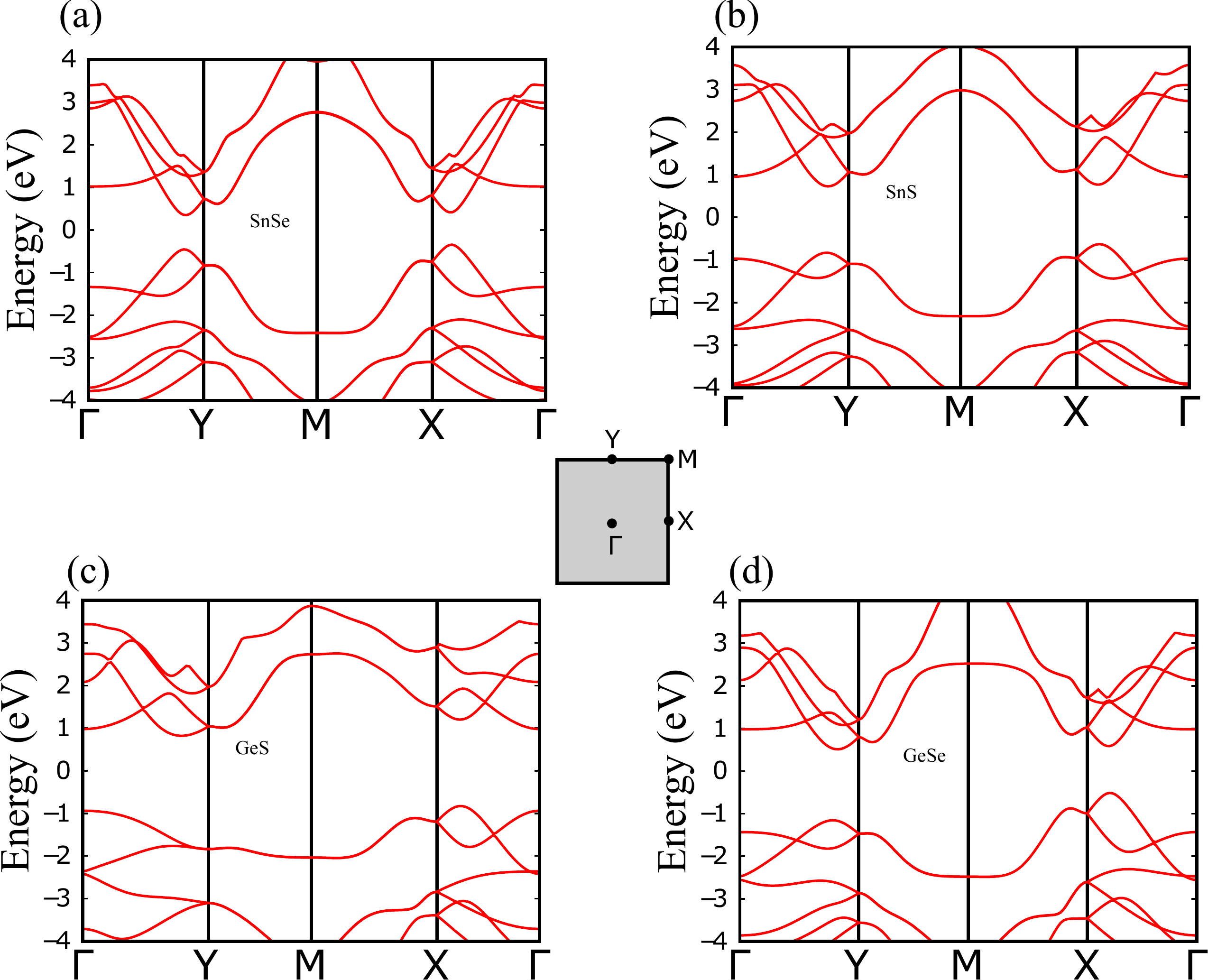}
	\caption{DFT-PBE calculated electronic band structure of (a) SnSe, (b) SnS, (c) GeS, and (d) GeSe monolayers. The shaded rectangular region represents the Brillouin zone. \label{fig:BS_all}}
		\end{minipage}
\end{figure*}
	\begin{table}[t]
	\caption{ DFT-PBE calculated band gap and lattice parameters of MX monolayers. \label{tab:t1}} 
	\centering
	\begin{tabular}{c c c c c c}
		\hline
		Material:    & SnSe& SnS & GeSe & GeS \\
		\hline
		$E_g$(eV) & 0.70 & 1.35 & 1.02 & 1.65 \\
		$|\bf a| $ (\AA)  & 4.31 & 4.20 & 4.27 & 4.47 \\
		$|\bf b|$ (\AA)  & 4.24 & 4.06 & 3.92 & 3.62 \\
		\hline
	\end{tabular} \\
\end{table}
\begin{figure*}[h]  
	\includegraphics[width=1\linewidth]{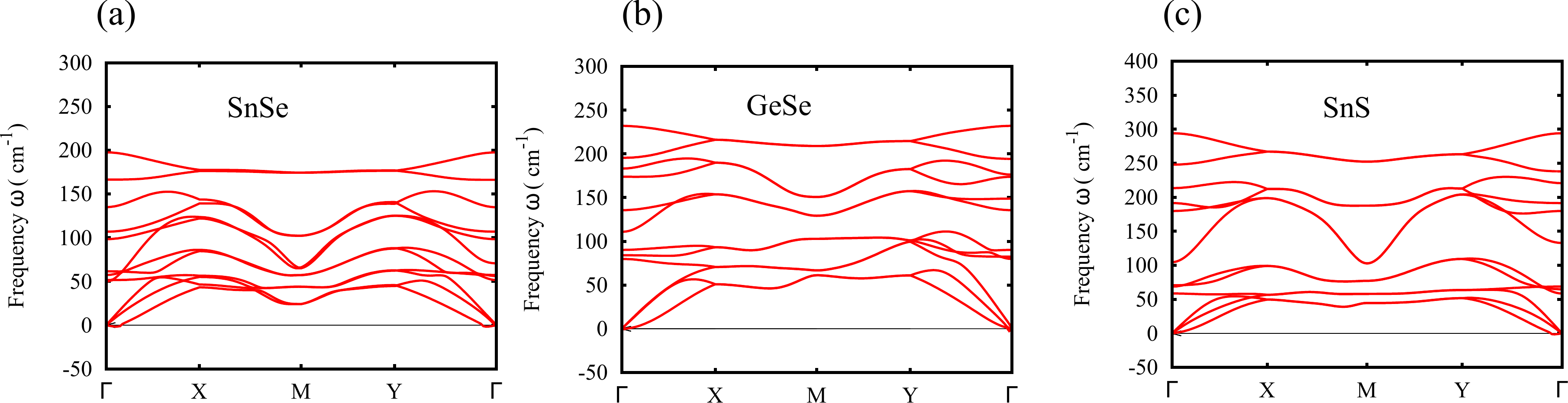}
	\caption{\textcolor{red}{Phonon band structure of a) SnSe, b) GeSe, and c) SnS monolayer without strain.} }
	\label{fig:strained_w}
\end{figure*}
\begin{figure*}[h]  
	\includegraphics[width=1\linewidth]{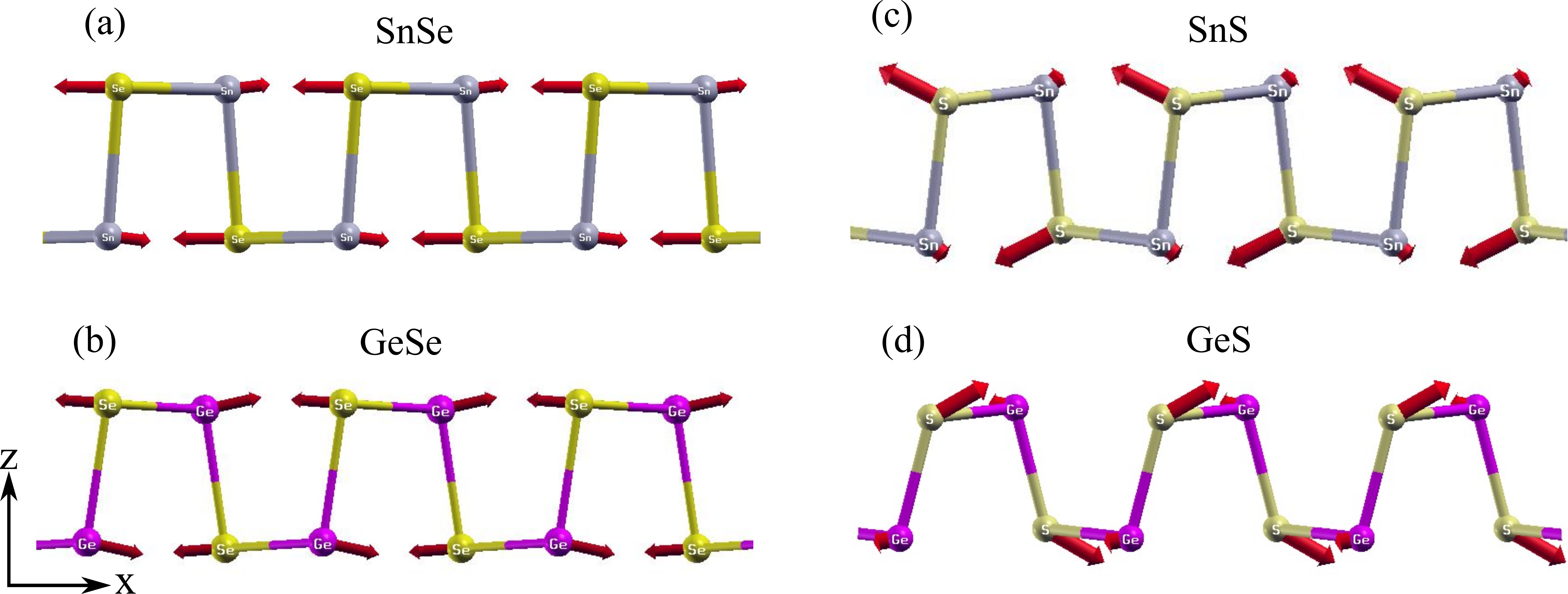}
	\caption{\textcolor{red}{Polar vibrational modes responsible for spontaneous polarization in (a) SnSe, (b) GeSe, (c) SnS, and (d) GeS monolayer at $\Gamma$ point.} }
	\label{fig:strained_w}
\end{figure*}
\begin{figure*}[b]  
	\includegraphics[width=1\linewidth]{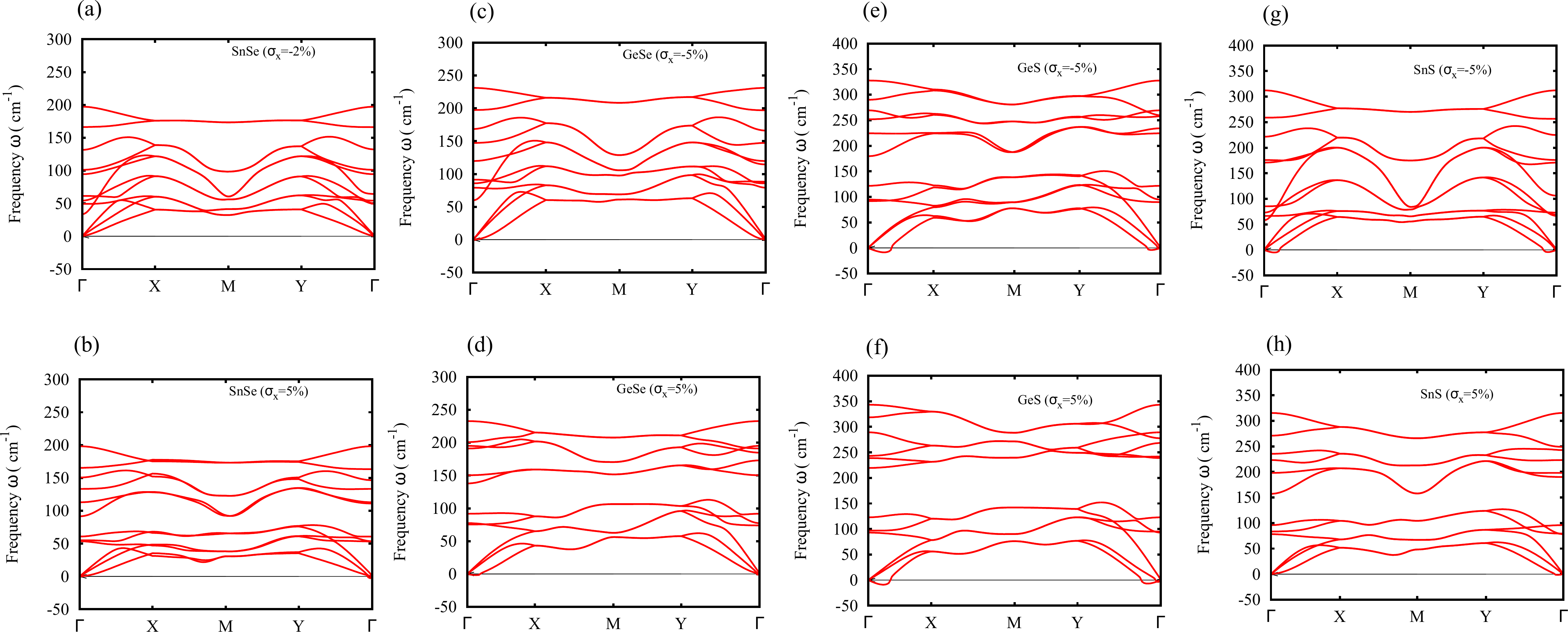}
	\caption{\textcolor{red}{Phonon band structures of considered monolayers at two extreme uniaxial strains at $\Gamma$ point. We have calculated the phonon dispersion at $\sigma_x=-2\%$  in the case of SnSe because the double well structure in the free energy vanishes at a compressive strain of 2\%.}  }
	\label{fig:strained_phonon}
\end{figure*}
\begin{figure*}   
	\includegraphics[width=\linewidth]{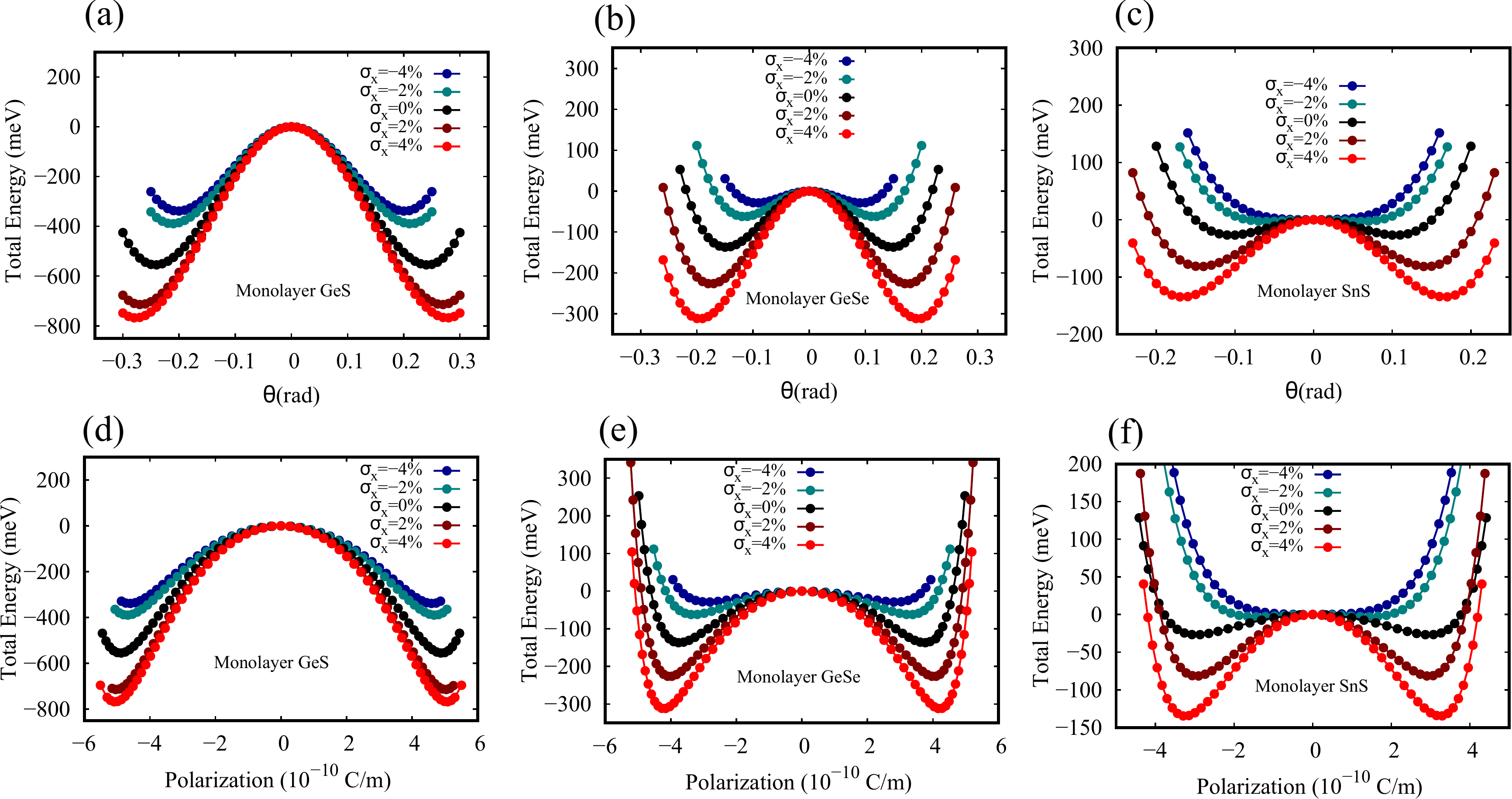}
	\caption{Strain induced modification in the double-well potential of GeS, GeSe, and SnS monolayers as a function of (a)-(c) $\theta$ , (d)-(f) polarization. }
	\label{fig:strained_w}
\end{figure*}
\begin{figure}[t]
	\centering	
	\includegraphics[width=\linewidth]{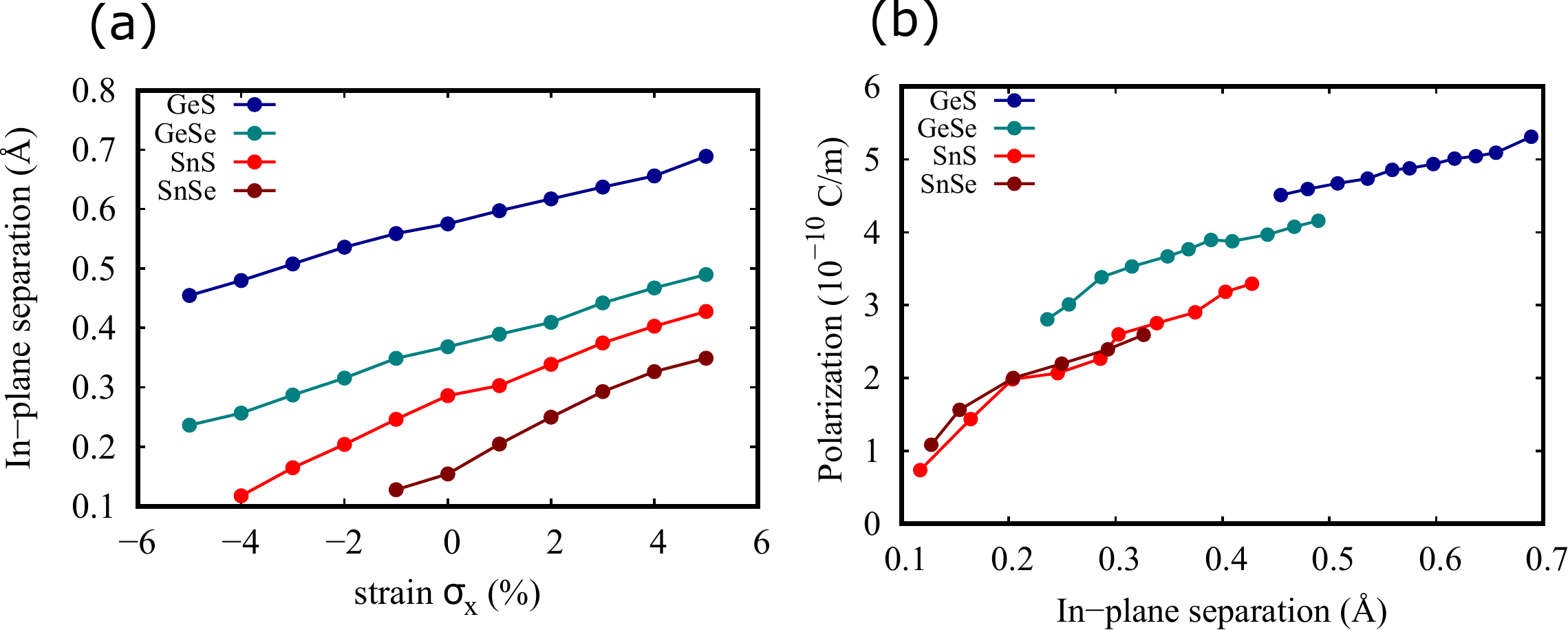}
	\caption{ (a) The change in the in-plane separation of MX (M=Ge, Sn;	X=S, Se) atoms as a function of strain $\sigma_x$, and (b) Polarization as a function of in-plane separation of MX atoms in all four monolayers.\label{fig:sup}}
\end{figure}
\begin{figure}[t]
	\centering	
	\includegraphics[width=\linewidth]{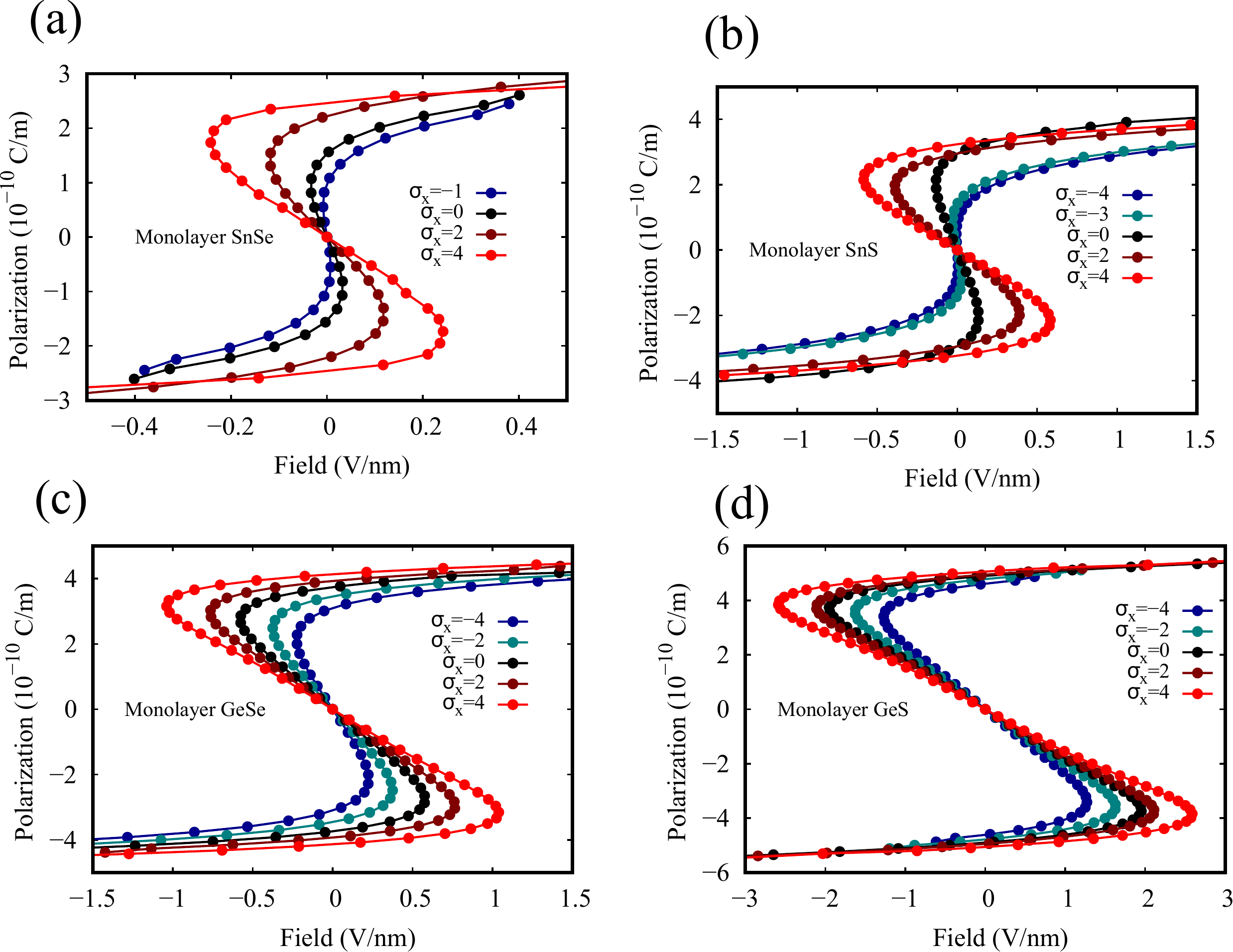}
	\caption{The strain dependence of polarization-field curve (or S-curve) for (a) SnSe, (b) SnS, (c) GeSe, and (d) GeS, obtained using the data of Figure~\ref{fig:strained_w}.\label{fig:sup}}
\end{figure}
\begin{figure}[h]
	\centering	
	\includegraphics[width=0.5\linewidth]{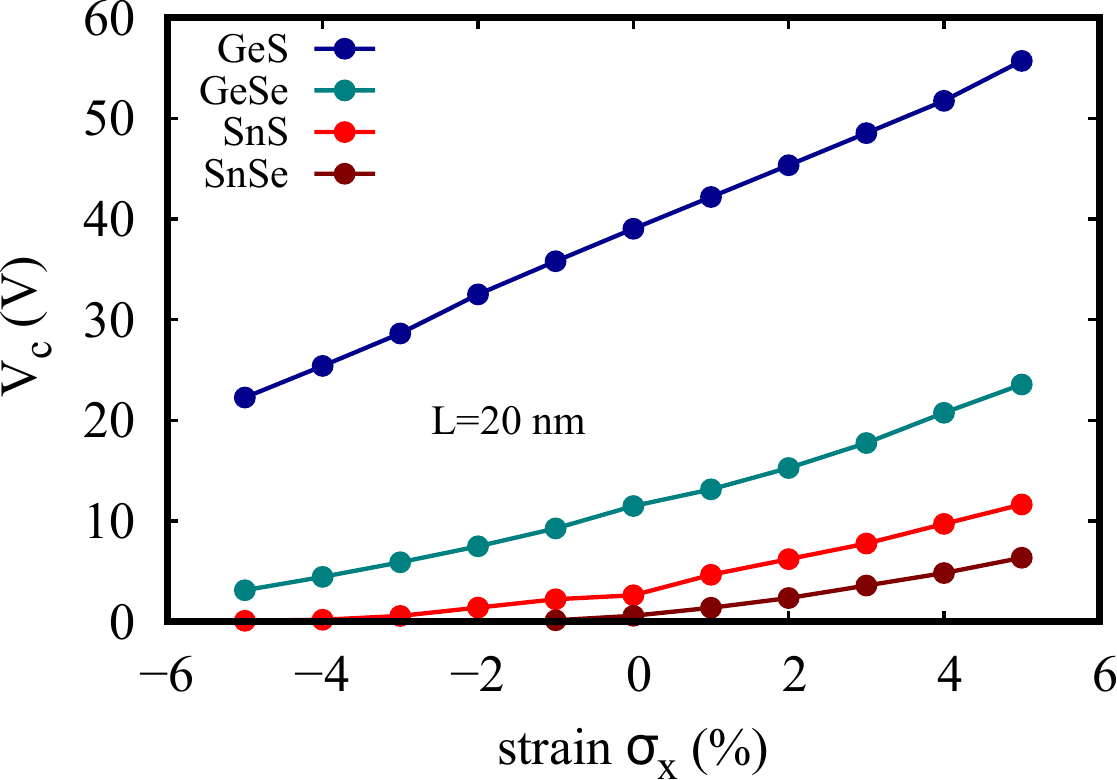}
	\caption{ The strain dependence of the coercive voltage V$_C$ for SnSe, SnS, GeSe, and GeS monolayers, calculated using V$_C=E_c\times L$.\label{fig:sup}}
\end{figure}
\begin{table}[h]
	\caption{Carrier effective mass and other relevant parameters of pristine monochalcogenides needed in TER calculation. \label{tab:t2}} 
	\centering
	\begin{tabular}{c c c c c}
		\hline
		Material:  & $m^{*}/m_{0}$ & $\epsilon_{FE}/\epsilon_{0}$ & $U_{0}(eV)$ & $\delta_{L,R}(\AA)$  \\
			\hline
		SnSe  & 0.12 & 7.69 & 0.35 & 4.41,2.20 \\
		SnS & 0.17 & 4.47 & 0.63 & 3.25,1.62 \\
		GeSe  & 0.13  & 3.18 & 0.51 & 2.66,1.33\\
		GeS & 0.21 & 3.14 & 0.82 & 2.65,1.32 \\
		\hline
	\end{tabular} \\
\end{table}